\documentclass[aps, prc, floatfix, nofootinbib, superscriptaddress, twocolumn]{revtex4-1}

\usepackage{latexsym}
\usepackage{amsmath}
\usepackage{amssymb}
\usepackage{amsfonts}

\usepackage[mathscr,scaled=1.15]{urwchancal}
\DeclareFontFamily{OT1}{pzc}{}
\DeclareFontShape{OT1}{pzc}{m}{it}%
{<-> s * [1.15] pzcmi7t}{}
\DeclareMathAlphabet{\mathpzc}{OT1}{pzc}{m}{it}

\usepackage{color}

\usepackage{supertabular}
\usepackage{placeins}
\usepackage{epsfig}
\usepackage{graphicx}
\usepackage{booktabs}

\definecolor{purple}{rgb}{0.5,0,0.5}
\definecolor{blue}{rgb}{0.0,0,0.9}
\definecolor{prdblue}{rgb}{0.133,0.118,0.498}
\usepackage[colorlinks=true, pdfstartview=FitV, linkcolor=prdblue, citecolor= prdblue, urlcolor=prdblue]{hyperref}

\begin{document}

\title{$\Upsilon(5S)$ in the unquenched quark model}

\author{Xiaoyun Chen}
\email[]{xychen@jit.edu.cn}
\affiliation{College of Science, Jinling Institute of Technology, Nanjing 211169, P. R. China}

\author{Yue Tan}
\email[]{tanyue@ycit.edu.cn}
\affiliation{Department of Physics, Yancheng Institute of Technology, Yancheng 224000, P. R. China}

\author{Yuan Chen}
\email[]{cycy@jit.edu.cn}
\affiliation{College of Science, Jinling Institute of Technology, Nanjing 211169, P. R. China}

\author{Xuejie Liu}
\email[]{1830592517@qq.com}
\affiliation{School of Physics, Henan Normal University, Xinxiang 453007, P. R. China}

\author{Jialun Ping}
\email[]{jlping@njnu.edu.cn}
\affiliation{College of Physics and Technology, Nanjing Normal University, Nanjing 211169, P. R. China}

\begin{abstract}
The observation of the $\Upsilon(10753)$ state by Belle II Collaboration has sparked significant interest in the theoretical understanding of such states within the context of hadron physics. Considering the similar mass and the decay with, as well as the same quantum numbers $J^{PC}=1^{--}$ with the $\Upsilon(10860)$ state, which is referred to be the $\Upsilon(5S)$ in PDG, in this work, we try to calculate the mass of the $\Upsilon(5S)$ state. The model used to predict the high-energy spectrum of these states generally involves a constituent quark model, which can describe a variety of properties of hadrons containing heavy quarks. In the framework of the unquenched quark model, a coupled-channel calculation is  employed to explore the effect of open-bottom meson-meson thresholds on the $\Upsilon(10860)$ state. The hypothesis is that coupled-channel effects could be large enough to create new dynamically generated states, thus potentially explaining the nature of the $\Upsilon(10860)$ state, as well as whether the $\Upsilon(10860)$ and $\Upsilon(10753)$ is the same state. The results indicate that unquenched effects play a crucial role in explaining the $\Upsilon(10860)$ state, providing a plausible mechanism for its formation. Besides in our calculations, the $\Upsilon(10860)$ and $\Upsilon(10753)$ may be two different states.
\end{abstract}

\maketitle

%%%%%%%%%%%%%%%%%%%%%%%%%%%%%%%%%%%%%%%%%%%%%%%%%%%%%%%%%%%%%%%%%%%%%%%%%%%%%%%%%%%%%%%%%%%%%%%%%%%%%%%%%%%%%%%%%%%%%%%

\section{Introduction} \label{introduction}
The study of heavy quarkonium spectroscopy, particularly the observation of higher states above the open heavy-flavor thresholds, offers a fascinating opportunity to deepen our understanding of the nonperturbative aspects of quantum chromodynamics (QCD). These studies provide critical insights into how quarks form different types of hadrons and are essential for advancing our knowledge of QCD's fundamental dynamics.

The charmonium and charmoniumlike $XYZ$ states above the $D^{(*)}\bar{D}^{(*)}$ thresholds have been extensively studied over the past two decades, with numerous experimental observations significantly enriching the field of hadron physics. These discoveries, as documented in numerous references (such as Refs. \cite{Liu:2013waa,Chen:2016qju,Liu:2019zoy,Guo:2017jvc,Brambilla:2019esw,Wang:2021aql}, have provided valuable insights into the mechanisms of hadron formation and the complex interactions that govern the behavior of quark-antiquark pairs in the meson sector.

However, when it comes to the bottomonium family, the situation is somewhat different. While only a few states within the bottomonium family have been observed thus far, the potential for new discoveries remains vast. The construction of a more comprehensive bottomonium family is still an open challenge and a subject of ongoing research. Observing and understanding the higher states in the bottomonium spectrum, particularly those above the open bottom-flavor thresholds, holds the promise of unlocking new information about QCD's nonperturbative regime.

Recently, the Belle II collaboration in 2022 provided additional confirmation of the $\Upsilon(10753)$ state by observing $\omega\chi_{bJ}(1P)$ signals at $\sqrt{s} = 10.745$ GeV \cite{Belle-II:2022xdi}. By combining the data from Belle II with the earlier results from Belle at $\sqrt{s} = 10.867$ GeV, they found that the energy dependence of the Born cross sections for $e^+e^- \rightarrow \omega\chi_{b_1,b_2}(1P)$ was consistent with the shape of the $\Upsilon(10753)$ resonance. The updated Breit-Wigner parameters for the state were:
\begin{align}
M&=(10753\pm6) ~\mathrm{MeV},  \\
\Gamma&=(36^{+18}_{-12})~ \mathrm{MeV},
\end{align}
and the quantum numbers $J^{PC} = 1^{--}$ were suggested for the state.

An enormous theoretical effort has followed the experimental discoveries. For example, in the relativistic flux tube model $\Upsilon(10753)$ is explained as a bottomonium state\cite{Chen:2019uzm}. Several other theoretical works also suggest that the $\Upsilon(10753)$ state could be a bottomonium-like state \cite{Liang:2019geg,ATLAS:2020yzc,Giron:2020qpb,vanBeveren:2020eis,Li:2021jjt,Bai:2022cfz}. Furthermore, there are various other interpretations for the $\Upsilon(10753)$ state, such as it being a hybrid state \cite{Brambilla:2019esw,TarrusCastella:2021pld}, a tetraquark state \cite{Ali:2019okl,Wang:2019veq,Bicudo:2020qhp,Bicudo:2022ihz}. 

The mass and the decay width of the new state $\Upsilon(10753)$ are both very close to the $\Upsilon(10860)$ state which is observed in 1985 by CLEO collaboration \cite{CLEO:1984vfn}. And in the experiment, the $\Upsilon(10860)$ is referred to the $\Upsilon(5S)$ in PDG (Particle Data Group) \cite{ParticleDataGroup:2024cfk} with the average mass 10885.2 MeV and decay with 37 MeV. And in the bottomonium region, the PDG states $\Upsilon(9460), \Upsilon(10023), \Upsilon(10355)$, and $\Upsilon(10579)$ are assigned to be $\Upsilon(1S)$ to $\Upsilon(4S)$ respectively.

All these states with the same quantum numbers $J^{PC} = 1^{--}$, as well as the new observed state $\Upsilon(10735)$ , whether the $\Upsilon(10735)$ signal is a normal $b\bar{b}$ state or not should be examined. 
It is significant to separate these exotic neutral states from conventional meson picture before treating them in other exotic pictures.

Obviously, it is not an easy task to establish the bottomonium spectrum completely because even many $b\bar{b}$ states below the $B\bar{B}$ threshold have not been discovered. However, the situation may be changed especially because of the running of Belle II. It is expected that more excited bottomonium
states will be detected in the near future. So it is time to investigate the spectrum of $b\bar{b}$ by diﬀerent approaches which incorporate the spirit of QCD.

In this work, within the framework of the constituent quark model \cite{Vijande:2004he}, satisfactorily describing a wide range of properties
of conventional hadrons containing heavy quarks, we primarily focus on studying the $\Upsilon(10860)$ state, since it has the almost mass and decay width with the observed new state $\Upsilon(10735)$. 
In this manuscript, the coupled-channels effects (Unquenched quark model) are adopted. The coupling between two-quark $b\bar{b}$ system and the four-quark system are investigated. The possible coupled-channels, such as $B\bar{B}$, $B\bar{B}^{*}, B^{*}\bar{B},B^{*}\bar{B}^{*}, B_s\bar{B}_s, B_s\bar{B}_s^*, B_s^*\bar{B}_s, B_s^*\bar{B}_s^*$, are considered in the unquenched quark model to shed light on the nature and structure of the $\Upsilon(5S)$. What's more, the Belle II Collaboration indicates that $\Upsilon(10860)$ and $\Upsilon(10753)$ states may not the same state \cite{Belle-II:2022xdi}. Through investigating the mass of the $\Upsilon(5S)$, it will help us better understand the experimental data.

In the calculations, we utilize a variational formalism based on a highly efficient numerical method known as the Gaussian Expansion Method (GEM) \cite{Hiyama:2003cu} to solve the bottomonium Hamiltonian. This Gaussian expansion approach enables us to compute effective meson-meson interactions from the original quark-(anti-)quark potentials in a simplified manner using the Resonating Group Method \cite{Wheeler:1937zza,Tang:1978zz}. Furthermore, in our framework, the coupling between the quark-antiquark and meson-meson sectors necessitates the creation of a light quark-antiquark pair. Consequently, the corresponding operator should be analogous to those used to describe open-flavor meson strong decays. For this, we adopt the $^3P_0$ transition operator as outlined in Ref. \cite{Micu:1968mk}. This theoretical formalism has the advantage of allowing the straightforward incorporation of coupling with all meson-meson partial waves and provides an efficient means to compute the probabilities associated with the different Fock components of the physical state.

Some previous studies have found that the creation of virtual quark pairs in hadronic systems leads to substantial mass shifts \cite{Barnes:2007xu}. These large mass shifts challenge the validity of the valence quark model in describing the ground state hadrons and raise concerns about the convergence of the Unquenched Quark Model (UQM). The convergence issue was pointed out by Ferretti and Santopinto, and they proposed that it could be addressed by considering only the contribution from the closest set of meson-meson intermediate states, while treating the contributions from other states as a global constant \cite{Ferretti:2018tco}. In our previous work \cite{Chen:2017mug,Chen:2024ukv,Chen:2023wfp}, we sought to resolve this problem by modifying the transition operator, specifically by introducing energy and separation damping factors. With the improved transition operator, the mass shifts of low-lying light mesons \cite{Chen:2017mug} and charmonium \cite{Chen:2023wfp} were significantly reduced. The proportion of the two-quark component increased to approximately 90\%, which in turn suppressed the influence of the four-quark components. This modification ensures the validity of the constituent valence quark model in describing the low-lying hadronic states. In this paper, we will continue to employ the modified model to calculate the $b\bar{b}$ system within the unquenched quark model, ensuring the continuity and validity of our calculations.

The paper is organized as follows. The theoretical framework is briefly presented in section II. Section III is mainly devoted to the analysis and discussion of our theoretical results. Finally, we summarize and draw some conclusions in Sec. IV.

%%%%%%%%%%%%%%%%%%%%%%%%%%%%%%%%%%%%%%%%%%%%%%%%%%%%%%%%%%%%%%%%%%%%%%%%%%%%%%%%%%%%%%%%%%%%%%%%%%%%%%%%%%%%%%%%%%%%%%%

\section{THEORETICAL FRAMEWORK}
\label{GEM and chiral quark model}
\subsection{Constituent quark model}
In the nonrelativistic quark model, we obtained the meson spectrum by solving
the Schr\"{o}dinger equation:
\begin{equation}
\label{Hamiltonian1} H \Psi_{M_I M_J}^{IJ} (1,2) =E^{IJ} \Psi_{M_I
M_J}^{IJ} (1,2)\,,
\end{equation}
where $1$, $2$ represents the quark and antiquark. $\Psi_{M_I M_J}^{IJ}(1,2)$ is the wave function of a meson composed of a quark and a antiquark with quantum numbers $IJ^{P}$ and reads,
\begin{align}
\nonumber
& \Psi_{M_I M_J}^{IJ}(1,2) \\
& =\sum_{\alpha}C_{\alpha} \left[
\psi_{l}(\mathbf{r})\chi_{s}(1,2)\right]^{J}_{M_J}
\omega^c(1,2)\phi^I_{M_I}(1,2), \label{PsiIJM}
\end{align}
where $\psi_{l}(\mathbf{r})$, $\chi_{s}(1,2)$,
$\omega^c(1,2)$, $\phi^I(1,2)$ are orbit, spin, color and flavor wave functions, respectively. $\alpha$ denotes the intermediate quantum numbers, $l,s$ and
possible flavor indices. In our calculations, the orbital wave functions is expanded using a set of Gaussians,
\begin{subequations}
\label{radialpart}
\begin{align}
\psi_{lm}(\mathbf{r}) & = \sum_{n=1}^{n_{\rm max}} c_{n}\psi^G_{nlm}(\mathbf{r}),\\
\psi^G_{nlm}(\mathbf{r}) & = N_{nl}r^{l}
e^{-\nu_{n}r^2}Y_{lm}(\hat{\mathbf{r}}),
\end{align}
\end{subequations}
with the Gaussian size parameters chosen according to the
following geometric progression
\begin{equation}\label{gaussiansize}
\nu_{n}=\frac{1}{r^2_n}, \quad r_n=r_1a^{n-1}, \quad
a=\left(\frac{r_{n_{\rm max}}}{r_1}\right)^{\frac{1}{n_{\rm
max}-1}}.
\end{equation}
This procedure enables optimization of the ranges using just a
small number of Gaussians.

At this point, the wave function in Eq.\,\eqref{PsiIJM} is expressed as follows:
\begin{align}
\nonumber
&\Psi_{M_I M_J}^{IJ}(1,2) \\
& =\sum_{n\alpha} C_{\alpha}c_n
 \left[ \psi^G_{nl}(\mathbf{r})\chi_{s}(1,2) \right]^{J}_{M_J}\omega^c(1,2)\phi^I_{M_I}(1,2).\label{Gauss1}
\end{align}
We employ the Rayleigh-Ritz variational principle for solving the Schr\"{o}dinger equation, which leads to a generalized eigenvalue problem due to the non-orthogonality of Gaussians
\begin{subequations}
\label{HEproblem}
\begin{align}
 \sum_{n^{\prime},\alpha^{\prime}} & (H_{n\alpha,n^{\prime}\alpha^{\prime}}^{IJ}
-E^{IJ} N_{n\alpha,n^{\prime}\alpha^{\prime}}^{IJ}) C_{n^{\prime}\alpha^{\prime}}^{IJ} = 0, \\
 &H_{n\alpha,n^{\prime}\alpha^{\prime}}^{IJ} =
  \langle\Phi^{IJ}_{M_I M_J,n\alpha}| H | \Phi^{IJ}_{M_I M_J,n^{\prime}\alpha^{\prime}}\rangle ,\\
 &N_{n\alpha,n^{\prime}\alpha^{\prime}}^{IJ}=
  \langle\Phi^{IJ}_{M_I M_J,n\alpha}|1| \Phi^{IJ}_{M_I M_J,n^{\prime}\alpha^{\prime}}\rangle,
\end{align}
\end{subequations}
with
$\Phi^{IJ}_{M_I M_J,n\alpha} =
[\psi^G_{nl}(\mathbf{r})\chi_{s}(1,2) ]^{J}_{M_J}
\omega^c(1,2)\phi^I_{M_I}(1,2)$,
$C_{n\alpha}^{IJ} = C_{\alpha}c_n$.

We obtain the mass of the four-quark system also by solving the Schr\"{o}dinger equation:
\begin{equation}
    H \, \Psi^{IJ}_{M_IM_J}(4q)=E^{IJ} \Psi^{IJ}_{M_IM_J}(4q),
\end{equation}
where $\Psi^{IJ}_{M_IM_J}(4q)$ is the wave function of the four-quark system, which can be constructed as follows. In our calculations, we only consider the color singlet-singlet
meson-meson picture for the four quark system. First, we write down the wave functions of two meson clusters,
\begin{subequations}
\label{Mesonfunctions}
\begin{align}
\nonumber
&    \Psi^{I_1J_1}_{M_{I_1}M_{J_1}}(1,2)=\sum_{\alpha_1 n_1} {\mathpzc C}^{\alpha_1}_{n_1} \\
    & \times  \left[ \psi^G_{n_1 l_1}(\mathbf{r}_{12})\chi_{s_1}(1,2)\right]^{J_1}_{M_{J_1}}
 \omega^{c_1}(1,2)\phi^{I_1}_{M_{I_1}}(1,2),   \\
&    \Psi^{I_2J_2}_{M_{I_2}M_{J_2}}(3,4)=\sum_{\alpha_2 n_2} {\mathpzc C}^{\alpha_2}_{n_2} \nonumber \\
    & \times \left[ \psi^G_{n_2 l_2}(\mathbf{r}_{34})\chi_{s_2}(3,4)\right]^{J_2}_{M_{J_2}}
    \omega^{c_2}(3,4)\phi^{I_2}_{M_{I_2}}(3,4),
\end{align}
\end{subequations}
then the total wave function of the four-quark state is:
\begin{align}
\label{4qfunctions}
& \Psi^{IJ}_{M_IM_J}(4q)  =  {\cal A} \sum_{L_r}\left[
\Psi^{I_1J_1}(1,2)\Psi^{I_2J_2}(3,4)
     \psi_{L_r}(\mathbf{r}_{1234})\right]^{IJ}_{M_IM_J}    \nonumber \\
\nonumber
  & =  \sum_{\alpha_1\,\alpha_2\,n_1\,n_2\,L_r}
  {\mathpzc C}^{\alpha_1}_{n_1} {\mathpzc C}^{\alpha_2}_{n_2} \bigg[ \left[\psi^G_{n_1 l_1}(\mathbf{r}_{12})\chi_{s_1}(1,2)\right]^{J_1} \nonumber \\
& \quad \times
            \left[\psi^G_{n_2 l_2}(\mathbf{r}_{34})\chi_{s_2}(3,4)\right]^{J_2}
             \psi_{L_r}(\mathbf{r}_{1234})\bigg]^{J}_{M_J} \nonumber \\
 &      \quad  \times \left[\omega^{c_1}(1,2)\omega^{c_2}(3,4)\right]^{[1]}
     \left[\phi^{I_1}(1,2)\phi^{I_2}(3,4)\right]^{I}_{M_I},
\end{align}
Here, ${\cal A}$ is the antisymmetrization operator, if all quarks
(antiquarks) are taken as identical particles, then
\begin{equation}
{\cal A}=\frac{1}{2}(1-P_{13}-P_{24}+P_{13}P_{24}).
\end{equation}
$\psi_{L_r}(\mathbf{r}_{1234})$ is the relative wave function between two clusters, which is also expanded in a set of Gaussians. $L_r$ is the relative orbital angular momentum.

The Hamiltonian of the chiral quark model for the four-quark system consists of three parts: quark rest mass, kinetic energy, potential energy (four-quark system is taken as an example):
\begin{align}
 H & = \sum_{i=1}^4 m_i  +\frac{p_{12}^2}{2\mu_{12}}+\frac{p_{34}^2}{2\mu_{34}}
  +\frac{p_{r}^2}{2\mu_{r}}  \quad  \nonumber \\
  & + \sum_{i<j=1}^4 \left( V_{\rm CON}^{C}(\boldsymbol{r}_{ij})+ V_{\rm OGE}^{C}(\boldsymbol{r}_{ij}) \right. \quad  \nonumber \\
  & \left. + V_{\rm CON}^{SO}(\boldsymbol{r}_{ij}) + V_{\rm OGE}^{SO}(\boldsymbol{r}_{ij}) +\sum_{\chi=\pi,K,\eta} V_{ij}^{\chi}
   +V_{ij}^{\sigma}\right).
\end{align}
Where $m_i$ is the constituent mass of $i$th quark (antiquark). $\frac{\bf{p^2_{ij}}}{2\mu_{ij}}~ (ij=12; 34)$ and $\frac{\bf{p^2_{r}}}{2\mu_{r}}$ represents the inner kinetic of two clusters
and the relative motion kinetic between two clusters, respectively, with
\begin{subequations}
\begin{align}
\bf{p}_{12}&=\frac{m_2\mathbf{p}_1-m_1\mathbf{p}_2}{m_1+m_2}, \\
\mathbf{p}_{34}&=\frac{m_4\mathbf{p}_3-m_3\mathbf{p}_4}{m_3+m_4},  \\
\mathbf{p}_{r}&= \frac{(m_3+m_4)\mathbf{p}_{12}-(m_1+m_2)\mathbf{p}_{34}}{m_1+m_2+m_3+m_4}, \\
\mu_{ij}&=\frac{m_im_j}{m_i+m_j}, \\
\mu_{r}&=\frac{(m_1+m_2)(m_3+m_4)}{m_1+m_2+m_3+m_4}.
\end{align}
\end{subequations}
$V_{\rm CON}^{C}$ and $V_{\rm OGE}^{C}$ are the central parts of the confinement and one-gluon-exchange. $V_{\rm CON}^{SO}$ and $V_{\rm OGE}^{SO}$ are the spin-orbit interaction potential energy.
In our calculations, a quadratic confining potential is adopted. For the mesons, the distance between $q$ and $\bar{q}$ is relatively small, so the difference between the linear potential
and the quadratic potential is very small by adjusting the confinement strengths. Both of them can conform to the linear Regge trajectories for $q\bar{q}$ mesons.
$V_{ij}^{\chi=\pi, K, \eta}$, and $\sigma$ exchange represents the one Goldstone boson exchange. Chiral symmetry suggests dividing quarks into two different sectors:
light quarks ($u$, $d$ and $s$) where the chiral symmetry is spontaneously broken and heavy quarks ($c$ and $b$) where the symmetry is explicitly broken.
The origin of the constituent quark mass can be traced back to the spontaneous breaking of chiral symmetry and consequently constituent quarks should interact through
the exchange of Goldstone bosons. The detailed derivation process can be found in the theoretical paper \cite{Machleidt:1987hj}. Here we only show the expressions of these potentials to save space.

The detailed expressions of the potentials are \cite{Valcarce:2005em}: {\allowdisplaybreaks
\begin{subequations}
\begin{align}
V_{\rm CON}^{C}(\boldsymbol{r}_{ij})&= ( -a_c r_{ij}^2-\Delta ) \boldsymbol{\lambda}_i^c \cdot \boldsymbol{\lambda}_j^c ,  \\
V_{\rm CON}^{\rm SO}(\boldsymbol{r}_{ij})&=\boldsymbol{\lambda}_i^c \cdot \boldsymbol{\lambda}_j^c \cdot \frac{-a_c}{2m_i^2m_j^2}\bigg\{ \bigg((m_i^2+m_j^2)(1-2a_s) \nonumber \\
&+4m_im_j(1-a_s)\bigg)(\boldsymbol{S_{+}} \cdot \boldsymbol{L})+(m_j^2-m_i^2) \nonumber \\
&(1-2a_s)(\boldsymbol{S_{-}} \cdot \boldsymbol{L})\bigg\}, \\
 V_{\rm OGE}^{C}(\boldsymbol{r}_{ij})&= \frac{\alpha_s}{4} \boldsymbol{\lambda}_i^c \cdot \boldsymbol{\lambda}_{j}^c
\left[\frac{1}{r_{ij}}-\frac{2\pi}{3m_im_j}\boldsymbol{\sigma}_i\cdot
\boldsymbol{\sigma}_j
  \delta(\boldsymbol{r}_{ij})\right],  \\
V_{\rm OGE}^{\rm SO}(\boldsymbol{r}_{ij})&=-\frac{1}{16}\cdot\frac{\alpha_s}{m_i^2m_j^2}\boldsymbol{\lambda}_i^c
\cdot \boldsymbol{\lambda}_j^c\big\{\frac{1}{r_{ij}^3}-\frac{e^{-r_{ij}/r_g(\mu)}}{r_{ij}^3}\cdot \nonumber \\
&(1+\frac{r_{ij}}{r_g(\mu)})\big\}\times \bigg\{\bigg((m_i+m_j)^2+2m_im_j\bigg)\nonumber \\
&(\boldsymbol{S_{+}} \cdot \boldsymbol{L})+(m_j^2-m_i^2)(\boldsymbol{S_{-}} \cdot \boldsymbol{L})\bigg\},  \\
\delta{(\boldsymbol{r}_{ij})} & =  \frac{e^{-r_{ij}/r_0(\mu_{ij})}}{4\pi r_{ij}r_0^2(\mu_{ij})}, \mathbf{S}_{\pm}=\mathbf{S}_1\pm \mathbf{S}_2,\\
V_{\pi}(\boldsymbol{r}_{ij})&= \frac{g_{ch}^2}{4\pi}\frac{m_{\pi}^2}{12m_im_j}
  \frac{\Lambda_{\pi}^2}{\Lambda_{\pi}^2-m_{\pi}^2}m_\pi v_{ij}^{\pi}
  \sum_{a=1}^3 \lambda_i^a \lambda_j^a,  \\
V_{K}(\boldsymbol{r}_{ij})&= \frac{g_{ch}^2}{4\pi}\frac{m_{K}^2}{12m_im_j}
  \frac{\Lambda_K^2}{\Lambda_K^2-m_{K}^2}m_K v_{ij}^{K}
  \sum_{a=4}^7 \lambda_i^a \lambda_j^a,   \\
\nonumber V_{\eta} (\boldsymbol{r}_{ij})& =
\frac{g_{ch}^2}{4\pi}\frac{m_{\eta}^2}{12m_im_j}
\frac{\Lambda_{\eta}^2}{\Lambda_{\eta}^2-m_{\eta}^2}m_{\eta}
v_{ij}^{\eta}  \\
 & \quad \times \left[\lambda_i^8 \lambda_j^8 \cos\theta_P
 - \lambda_i^0 \lambda_j^0 \sin \theta_P \right],   \\
v_{ij}^{\chi}(\boldsymbol{r}_{ij}) & =  \left[ Y(m_\chi r_{ij})-
\frac{\Lambda_{\chi}^3}{m_{\chi}^3}Y(\Lambda_{\chi} r_{ij})
\right]
\boldsymbol{\sigma}_i \cdot\boldsymbol{\sigma}_j,\\
V_{\sigma}(\boldsymbol{r}_{ij})&= -\frac{g_{ch}^2}{4\pi}
\frac{\Lambda_{\sigma}^2}{\Lambda_{\sigma}^2-m_{\sigma}^2}m_\sigma \nonumber \\
& \quad \times \left[
 Y(m_\sigma r_{ij})-\frac{\Lambda_{\sigma}}{m_\sigma}Y(\Lambda_{\sigma} r_{ij})\right]  ,
\end{align}
\end{subequations}}
\hspace*{-0.5\parindent}%
where $\mathbf{S}_1$ and $\mathbf{S}_2$ is the spin of the two meson clusters. $Y(x)  =   e^{-x}/x$; $r_0(\mu_{ij}) =s_0/\mu_{ij}$; $\boldsymbol{\sigma}$ are the $SU(2)$
Pauli matrices; $\boldsymbol{\lambda}$, $\boldsymbol{\lambda}^c$ are $SU(3)$ flavor, color Gell-Mann matrices, respectively. The form factor parameter $\Lambda_{\chi}$
$(\chi=\pi, K, \eta, \sigma)$ is introduced to remove the short-range contribution of Goldstone bosons exchanges. $g^2_{ch}/4\pi$ is the chiral coupling constant, determined from
the $\pi$-nucleon coupling; and $\alpha_s$ is an effective scale-dependent running coupling
\cite{Valcarce:2005em},
\begin{equation} \label{alphas}
\alpha_s(\mu_{ij})=\frac{\alpha_0}{\ln\left[(\mu_{ij}^2+\mu_0^2)/\Lambda_0^2\right]}.
\end{equation}
In our calculations, for the two-quark system, besides the central potential energy, the noncentral potential energy is also included. But in the four-quark system calculations,
we find that the influence of the noncentral potential energy on the mass shift of the state is tiny, so it is omitted.

Lastly, we show the model parameters in Table \ref{modelparameters}. All the parameters are determined by fitting the meson spectrum, from light meson to heavy mesons, taking into account only a quark-antiquark component. They are shown in Table \ref{modelparameters}. Using the model parameters, we calculated the mesons involved in the present work in Table \ref{BBmass}. Furthermore, we show the charmonium and bottomonium mesons spectrum in the constituent quark model, which are shown in Table \ref{ccbbmass}. It should be noted that, in our calculation, we did not use the formula for $\alpha_s$ in Eq. (\ref{alphas}). Instead, for different quarks, we obtained the values of $\alpha_s$ between different quarks (Table \ref{modelparameters}) through fitting experimental data.
\begin{table}[!t]
\begin{center}
\caption{ \label{modelparameters} Model parameters, determined by fitting the meson spectrum, leaving room for unquenching
contributions in the case of light-quark systems.}
\begin{tabular}{llr}
\hline\noalign{\smallskip}
Quark masses   &$m_u=m_d$     &490  \\
   (MeV)       &$m_s$         &511  \\
               &$m_c$         &1649 \\
               &$m_b$         &4978 \\
\hline
Goldstone bosons                &$m_{\pi}$        &0.70  \\
(fm$^{-1} \sim 200\,$MeV )      &$m_{\sigma}$     &3.42  \\
                                &$m_{\eta}$       &2.77  \\
                                &$m_{K}$          &2.51  \\
                                &$\Lambda_{\pi}=\Lambda_{\sigma}$     &3.5  \\
                                &$\Lambda_{\eta}=\Lambda_{K}$         &2.2  \\
                   \cline{2-3}
                               &$g_{ch}^2/(4\pi)$                     &0.54  \\
                               &$\theta_p(^\circ)$                    &-15 \\
\hline
Confinement                    &$a_c$ (MeV fm$^{-2}$)                 &98.0 \\
                               &$\Delta$ (MeV)                        &-18.1 \\
                               &$a_s$                                 &0.77 \\
\hline
OGE                            &$\alpha_{qq}$                         &1.34 \\
                               &$\alpha_{ss}$                         &0.92 \\
                               &$\alpha_{bb}$                         &0.43 \\
                               &$\alpha_{cc}$                         &0.56 \\
                               &$\alpha_{qs}$                         &1.15 \\
                               &$\alpha_{qb}$                         &0.75 \\
                               &$\alpha_{qc}$                         &0.85 \\
                               &$\alpha_{sc}$                         &0.75 \\
                               &$\alpha_{sb}$                         &0.65 \\
                               &$\alpha_{bc}$                         &0.49 \\
                               &$r_g$(MeV)                            &100.6\\
                               &$r_0$(MeV)                            &81.0 \\
\hline
\end{tabular}
\end{center}
\end{table}

\begin{table}[!t]
\begin{center}
\renewcommand{\arraystretch}{1.8}
\setlength{\tabcolsep}{1.5mm}
\caption{ \label{BBmass} The involved mesons in the present work.(unit of MeV)}
\begin{tabular}{ccccccc}
\hline\noalign{\smallskip}
Meson     &$B^+$   &$B^-$  &$B^{*+}$  &$B^{*-}$  &$B_s^0$  &$B_s^*$ \\
\hline
$I(J^P)$  &$\frac{1}{2}(0^-)$ &$\frac{1}{2}(0^-)$ &$\frac{1}{2}(1^-)$ &$\frac{1}{2}(1^-)$ &$0(0^-)$ &$0(1^-)$ \\
Mass      &5278.5  &5278.5 &5334.1    &5334.1    &5373.3   &5417.1\\
Exp.\cite{ParticleDataGroup:2024cfk}      &5279.4  &5279.4 &5324.7    &5324.7    &5366.9   &5415.4\\
\hline
\end{tabular}
\end{center}
\end{table}

\begin{table*}[!t]
\begin{center}
\setlength{\tabcolsep}{5mm}
\caption{ \label{ccbbmass} The mass spectrum of charmonium and bottomonium states using the model parameters in Table \ref{modelparameters}.(unit of MeV)}
\begin{tabular}{ccccccccc}
\hline\noalign{\smallskip}
\multicolumn{4}{c}{Charmonium States}  &&\multicolumn{4}{c}{Bottomonium States} \\
\cline{1-4}\cline{6-9} 
State              &$J^{PC}$   &Mass     &PDG\cite{ParticleDataGroup:2024cfk} &&State &$J^{PC}$   &Mass     &PDG\cite{ParticleDataGroup:2024cfk} \\
\hline
$\eta_c$           &$0^{-+}$   &2977.6   &2984.1 &&$\eta_b$           &$0^{-+}$   &9388.9   &9398.7 \\
$J/\psi$           &$1^{--}$   &3099.9   &3096.9 &&$\Upsilon(1S)$     &$1^{--}$   &9498.8   &9460.4\\
$\chi_{c_0}(1P)$   &$0^{++}$   &3438.8   &3414.7 &&$\chi_{b_0}(1P)$   &$0^{++}$   &9874.9   &9859.4\\
$\chi_{c_1}(1P)$   &$1^{++}$   &3471.3   &3510.7 &&$\chi_{b_1}(1P)$   &$1^{++}$   &9888.4   &9892.7\\
$h_c(1P)$          &$1^{+-}$   &3492.8   &3525.4 &&$h_b(1P)$          &$1^{+-}$   &9898.8   &9899.3\\
$\chi_{c_2}(1P)$   &$2^{++}$   &3524.5   &3556.2 &&$\chi_{b_2}(1P)$   &$2^{++}$   &9910.4   &9912.2\\
$\eta_c(2S)$       &$0^{-+}$   &3602.3   &3637.7 &&$\Upsilon(2S)$     &$1^{--}$   &9968.4   &10023.4\\ 
$\psi(2S)$         &$1^{--}$   &3666.1   &3686.1 &&$\Upsilon(1D)$     &$2^{--}$   &10098.5  &10163.7\\ 
$\psi(3770)$       &$1^{--}$   &4155.8   &3773.7 &&$\chi_{b_0}(2P)$   &$0^{++}$   &10176.4  &10232.5 \\
$\chi_{c_0}(3915)$ &$0^{++}$   &3923.1   &3922.1 &&$\chi_{b_1}(2P)$   &$1^{++}$   &10188.2  &10255.5\\ 
$\chi_{c_1}(3872)$ &$1^{++}$   &3951.8   &3871.6 &&$h_b(2P)$          &$1^{+-}$   &10197.2  &10259.8\\ 
$\chi_{c_2}(3930)$ &$2^{++}$   &4001.4   &3922.5 &&$\chi_{b_2}(2P)$   &$2^{++}$   &10208.2  &10268.6\\ 
$\psi(4040)$       &$1^{--}$   &4634.7   &4040.0 &&$\Upsilon(3S)$     &$1^{--}$   &10281.8  &10355.1\\ 
                   &           &         &       &&$\chi_{b_1}(3P)$   &$1^{++}$   &10468.6  &10513.4\\ 
                   &           &         &       &&$\chi_{b_2}(3P)$   &$2^{++}$   &10488.2  &10524.0 \\
                   &           &         &       &&$\Upsilon(4S)$     &$1^{--}$   &10579.6  &10579.4 \\
                   &           &         &       &&$\Upsilon(10860)$  &$1^{--}$   &10927.6  &10885.2\\ 
                   &           &         &       &&$\Upsilon(11020)$  &$1^{--}$   &11167.8  &11000.0\\ 
\hline
\end{tabular}
\end{center}
\end{table*}

As shown in the Table \ref{BBmass}, our theoretical results for the $B^{(*)}$ and $B_s^{(*)}$ mesons are in very good agreement with the experimental values.
From the Table \ref{ccbbmass}, we observe that our model provides a good description of the ground states in the charmonium sector, such as the $\eta_c$, $J/\psi$, $\chi_{c_J}(1P) (J=0,1,2)$ and so on. For bottomonium states, our model also yields reasonably good theoretical results for the ground-state mesons. For example, the masses of the $\eta_b$, $\Upsilon(1S)$, $\chi_{b_J}(1P)(J=0,1,2)$ are well reproduced within our framework.

%%%%%%%%%%%%%%%%%%%%%%%%%%%%%%%%%%%%%%%%%%%%%%%%%%%%%%%%%%%%%%%%%%%%%%%%%%%%%%%%%%%%%%%%%%%%%%%%%%%%%%%%%%%%%%%%%%%%%%%
\subsection{The transition operator}
\label{sec3P0}
The $^3P_0$ quark-pair creation model \cite{Micu:1968mk,
LeYaouanc:1972vsx, LeYaouanc:1973ldf} has been widely applied to
OZI rule allowed two-body strong decays of hadrons
\cite{Roberts:1992js, Capstick:1993kb,
Page:1995rh, Ackleh:1996yt, Segovia:2012cd}. If the quark and antiquark in the original meson are labeled by 1, 2, and the quark and antiquark ($u\bar{u}$, $d\bar{d}$,
$s\bar{s}$) generated in the vacuum are numbered as 3, 4, the transition operator of the $^3P_0$ model reads:
\begin{align} \label{T0}
T_0 & =-3\, \gamma \sum_m\langle 1m1(-m)|00\rangle\int
d\mathbf{p}_3d\mathbf{p}_4\delta^3(\mathbf{p}_3+\mathbf{p}_4)\nonumber\\
& \quad \times{\cal{Y}}^m_1(\frac{\mathbf{p}_3-\mathbf{p}_4}{2})
\chi^{34}_{1-m}\phi^{34}_0\omega^{34}_0b^\dagger_2(\mathbf{p}_3)d^\dagger_3(\mathbf{p}_4),
\end{align}
where, $\chi^{34}_{1-m},\phi^{34}_0,\omega^{34}_0$ are spin, flavor and color wave functions of the created quark pair, respectively. ${\cal{Y}}^{m}_{1}(\frac{\mathbf{p}_3-\mathbf{p}_4}{2})$ = $pY^m_1(\hat{\mathbf{p}})$ is the solid spherical harmonics. $\gamma$ describes the probability for creating
a quark-antiquark pair with momenta $\mathbf{p}_3$ and $\mathbf{p}_4$ from the vacuum. It is normally determined by fitting the strong decay widths of hadrons.
This yields $\gamma=6.95$ for $u\bar{u}$ and $d\bar{d}$ pair creation, and $\gamma=6.95/\sqrt{3}$ for $s\bar{s}$ pair creation \cite{LeYaouanc:1977gm}.

To reduce the mass shift due to the coupled-channel effects, the transition operator in Eq. (\ref{T0}) should be modified. In Ref. \cite{Chen:2017mug}, two suppression factors are introduced, namely energy damping factor and distance damping factor.
The first factor is $\exp[-r^2/(4f^2)]$ ($\exp[-f^2 p^2]$ in momentum space), here $\mathbf{r}=\mathbf{r_3}-\mathbf{r_4}$ is the distance between the quark and antiquark created in the vacuum, considering the effect of quark-antiquark energy created in the vacuum and it suppresses the contribution from meson-meson states with high energy. What's more, when the distance between the bare meson and a pair of mesons becomes smaller, the energy of tetraquark will increase, and the momentum of the created quark (antiquark) will be large. At this point, the energy damping factor $\exp[-f^2 p^2]$ comes into play, thus the mass shift of the mesons is still suppressed and the convergence is guaranteed.
The second factor is $\exp[-R_{AV}^2/R_0^2]$, which takes into account the effect that the created
quark-antiquark pair should not be far away from the source meson. Here, $R_{AV}$ represents the distance between the created quark-antiquark pair and the source meson. It reads,
\begin{subequations}
\begin{align}
\mathbf{R_{AV}}&= \mathbf{R_A}-\mathbf{R_V};\\
\mathbf{R_A}&=\frac{m_1\mathbf{r_1}+m_2\mathbf{r_2}}{m_1+m_2}; \\
\mathbf{R_V}&= \frac{m_3\mathbf{r_3}+m_4\mathbf{r_4}}{m_3+m_4}=\frac{\mathbf{r_3}+\mathbf{r_4}}{2}~~ (m_3=m_4).
\end{align}
\end{subequations}
So the modified transition operator takes
\begin{align} \label{T1}
T_1&= -3\gamma\sum_{m}\langle 1m1(-m)|00\rangle\int
d\mathbf{r_3}d\mathbf{r_4}(\frac{1}{2\pi})^{\frac{3}{2}}ir2^{-\frac{5}{2}}f^{-5}
\nonumber \\
 & Y_{1m}(\hat{\mathbf{r}})
 {\rm e}^{-\frac{\mathbf{r}^2}{4f^2}}
 {\rm e}^{-\frac{R_{AV}^2}{R_0^2}}\chi_{1-m}^{34}\phi_{0}^{34}
 \omega_{0}^{34}b_3^{\dagger}(\mathbf{r_3})d_4^{\dagger}(\mathbf{r_4}),
\end{align}
By fitting the decay width of $\rho \rightarrow \pi\pi$, and with the requirement that the mass shift is around the $10\%$ of the bare mass, the parameters $f$, $R_0$ and $\gamma$ were fixed \cite{Chen:2017mug,Chen:2024ukv,Chen:2023wfp},
\begin{equation}\label{para}
\gamma = 32.2,\; \quad f=0.5\,\mbox{fm},\; \quad R_0=1\,\mbox{fm}.
\end{equation}

In this work, when we study the effects of the coupled-channel in the $b\bar{b}$ system, we will continue using this set of $^3P_0$ parameters.

%%%%%%%%%%%%%%%%%%%%%%%%%%%%%%%%%%%%%%%%%%%%%%%%%%%%%%%%%%%%%%%%%%%%%%%%%%%%%%%%%%%%%%%%%%%%%%%%%%%%%%%%%%%%%%%%%%%%%%%

\section{Numerical Results}
\label{Numerical Results}
In UQM, we obtain the eigenvalues of systems (quark-antiquark plus four-quark components) by solving the Schr\"{o}dinger equation,
\begin{eqnarray}
H\Psi=E\Psi ,
\end{eqnarray}
where $\Psi$ and $H$ is the wave function and the Hamiltonian of the system, respectively. It reads,
\begin{eqnarray}
\Psi=c_1\Psi_{2q}+c_2\Psi_{4q} ~,\\
 H=H_{2q}+H_{4q}+T ~.
\end{eqnarray}
The term $H_{2q}$ only acts on the wave function of two-quark system, $\Psi_{2q}$, and the  $H_{4q}$ only acts on the wave function of four-quark system, $\Psi_{4q}$. The transition operator $T$ is responsible for mixing the quark-antiquark and four-quark components.

In this way, we can get the matrix elements of the Hamiltonian,
\begin{align}
\langle\Psi| & H|\Psi\rangle = \langle
c_1\Psi_{2q}+c_2\Psi_{4q}|H|c_1\Psi_{2q}+c_2\Psi_{4q}\rangle
\nonumber \\
&=c_1^2\langle\Psi_{2q}|H_{2q}|\Psi_{2q}\rangle+c_2^2\langle\Psi_{4q}|H_{4q}|\Psi_{4q}\rangle
\nonumber \\
&\quad+c_1c_2^*\langle\Psi_{4q}|T|\Psi_{2q}\rangle+c_1^*c_2\langle\Psi_{2q}|T^{\dagger}|\Psi_{4q}\rangle,
\end{align}
and the block-matrix structure for the Hamiltonian and overlap
takes,
\begin{equation}
(H)=\left[\begin{array}{cc} (H_{2q}) & (H_{24})\\
(H_{42}) & (H_{4q})
\end{array}
\right],
(N)=\left[\begin{array}{cc} (N_{2q}) & (0)\\
(0) & (N_{4q})
\end{array}
\right] \,,
\end{equation}
with
{\allowdisplaybreaks
\begin{subequations}
\begin{align}
 (H_{2q})&=\langle\Psi_{2q}|H_{2q}|\Psi_{2q}\rangle, \\
 (H_{24})&=\langle\Psi_{4q}|T|\Psi_{2q}\rangle, \\
 (H_{4q})&=\langle\Psi_{4q}|H_{4q}|\Psi_{4q}\rangle,\\
(N_{2q})&=\langle\Psi_{2q}|1|\Psi_{2q}\rangle, \\
(N_{4q})&=\langle\Psi_{4q}|1|\Psi_{4q}\rangle.
\end{align}
\end{subequations}
}
Where $(H_{2q})$ and $(H_{4q})$ is the matrix for the pure two-quark system and pure four-quark system. $(N_{2q})$ and $(N_{4q})$ is their respective overlap matrix. $(H_{24})$ is the coupling matrix of two-quark system and four-quark system.

Finally the eigenvalues ($E_n$) and eigenvectors ($C_n$) of the system are obtained by solving the generalized eigen-problem,
\begin{eqnarray}
\Big[
\begin{array}{c}
(H)-E_n(N)
\end{array}
\Big] \Big[
\begin{array}{c} C_n
\end{array}
\Big]=0. \label{geig}
\end{eqnarray}

For the $\Upsilon(10860)$, the quantum numbers is $I^G(J^{PC})=0^-(1^{--})$. In the framework of two-quark system, the orbital angular momentum is 0, the spin angular momentum is 1, and the total angular momentum is 1. In the framework of four-quark system, for two color-singlet mesons, we assume that each meson has internal orbital angular momentum $l_{1}=l_{2}=0$ in Eq. (\ref{Mesonfunctions}). The relative orbital angular momentum between the two mesons is taken to be $L_{r}=1$ in Eq. (\ref{4qfunctions}). The total spin angular momentum of the system can then take values $s_1 \otimes s_2=S(S=0,1,2)$. The total angular momentum is fixed at $J=1$ in Eq. (\ref{4qfunctions}), with its third component $M_{J}=1$. The choice of $M_{J}=1$ does not affect the final theoretical results. What's more, the color wave functions of both mesons are taken to be color-singlet states.

In Table \ref{channels}, we list the possible tetraquark states that can couple to the two-quark configurations. Here, $s_1$ and $s_2$ denote the spins of the two mesons, and $S$ is the total angular momentum of the four quark system with $s_1 \otimes s_2 =S$. $I_1$ and $I_2$ represent their isospins, and $I$ is the total isospin with $I_1 \otimes I_2 = I$. $c_1$ and $c_2$ refer to the color degrees of freedom of the quark pairs. The case $c_1=c_2=1$ corresponds to color-singlet mesons. The last column indicates the possible coupling channels.

\begin{table}[!t]
\begin{center}
\setlength{\tabcolsep}{3mm}
\caption{ \label{channels} The possible tetraquark states that can couple to the two-quark configurations.}
\begin{tabular}{ccccccccc}
\hline\noalign{\smallskip}
$s_1$   &$s_2$  &$S$   &$I_1$           &$I_2$          &$I$   &$c_1$   &$c_2$  &channel \\
\hline
0       &0      &0     &$\frac{1}{2}$   &$\frac{1}{2}$  &0     &1       &1      &$B\bar{B}$ \\
1       &1      &0     &$\frac{1}{2}$   &$\frac{1}{2}$  &0     &1       &1      &$B^{*}\bar{B}^{*}$\\
0       &1      &1     &$\frac{1}{2}$   &$\frac{1}{2}$  &0     &1       &1      &$B\bar{B}^{*}$\\
1       &0      &1     &$\frac{1}{2}$   &$\frac{1}{2}$  &0     &1       &1      &$B^{*}\bar{B}$\\
1       &1      &1     &$\frac{1}{2}$   &$\frac{1}{2}$  &0     &1       &1      &$B^{*}\bar{B}^{*}$\\
1       &1      &2     &$\frac{1}{2}$   &$\frac{1}{2}$  &0     &1       &1      &$B^{*}\bar{B}^{*}$\\
0       &0      &0     &0               &0              &0     &1       &1      &$B_s\bar{B}_s$ \\
1       &1      &0     &0               &0              &0     &1       &1      &$B_s^*\bar{B}_s^*$ \\
0       &1      &1     &0               &0              &0     &1       &1      &$B_s\bar{B}_s^*$ \\
1       &0      &1     &0               &0              &0     &1       &1      &$B_s^*\bar{B}_s$ \\
1       &1      &1     &0               &0              &0     &1       &1      &$B_s^*\bar{B}_s^*$ \\
1       &1      &2     &0               &0              &0     &1       &1      &$B_s^*\bar{B}_s^*$ \\
\hline
\end{tabular}
\end{center}
\end{table}

For the bottomonium states, there are about six $\Upsilon(nS)$ with quantum numbers $J^{PC}=1^{--}$ in PDG \cite{ParticleDataGroup:2024cfk}. In this paper, we aim to investigate the mass shift of the $\Upsilon(5S)$ state within the unquenched quark model, examining whether the inclusion of coupled-channel effects can lower its mass to approximately the $\Upsilon(10753)$ state. The parameters of the quark model play a crucial role in these calculations. Table \ref{modelparameters} presents the model parameters used in this work, which were determined by fitting experimental data for the ground states of both light and heavy quark mesons. As a result, the model parameters provides a good description of the ground states of light and heavy mesons but performs poorly for highly excited states. Based on these model parameters, we study the unquenched effects in high-lying excitations and compare our results with experimental data, thereby improving our understanding of the observations.

As shown in the Table \ref{channels}, there are 12 possible coupling channels. Within the unquenched quark model framework, we will compute the coupling effects between each of these meson-meson channels and the $b\bar{b}$ two-quark state $\Upsilon(5S)$. In our two-quark model calculation, the mass of $\Upsilon(5S)$ is found to be 10927.6 MeV, seen in Table \ref{ccbbmass}. After considering the coupled-channel effects, the resulting mass shifts of the mesons are presented in Table \ref{results}. 

It is important to note that in the calculation, the mass shift varies with spatial configuration. Therefore, obtaining a stable mass shift is crucial. Only when the mass of the scattering state in the tetraquark sector is very close to that of the two-quark state (10927.6 MeV) does the coupling become strongest. Besides, when we adjust the range of the Gaussian to bring the energy of another scattering state close to that of 10927.6 MeV, the mass shift induced by the coupling between the two-quark and four-quark components remains the same. Lastly, largest and most stable mass shift is obtained. 

\begin{table}[!t]
\begin{center}
\setlength{\tabcolsep}{5.4mm}
\caption{ \label{results} The mass shifts of the possible coupled-channels for the $\Upsilon(10860)$ state in unit of MeV.}
\begin{tabular}{ccccc}
\hline\noalign{\smallskip}
$s_1$   &$s_2$  &$S$   &channel   &Mass Shift  \\
\hline
0       &0      &0     &$B\bar{B}$         &-4.87\\
1       &1      &0     &$B^{*}\bar{B}^{*}$ &-2.39\\
0       &1      &1     &$B\bar{B}^{*}$     &-1.96\\
1       &0      &1     &$B^{*}\bar{B}$     &-1.96  \\
1       &1      &1     &$B^{*}\bar{B}^{*}$ &-0.0\\
1       &1      &2     &$B^{*}\bar{B}^{*}$ &+8.0\\
0       &0      &0     &$B_s\bar{B}_s$     &-2.99 \\
1       &1      &0     &$B_s^*\bar{B}_s^*$ &-2.4\\
0       &1      &1     &$B_s\bar{B}_s^*$   &-8.22\\
1       &0      &1     &$B_s^*\bar{B}_s$   &-8.22\\
1       &1      &1     &$B_s^*\bar{B}_s^*$ &-0.0\\
1       &1      &2     &$B_s^*\bar{B}_s^*$ &-6.4\\
\hline
\multicolumn{5}{c}{Total mass shift: -31.4 MeV} \\
\multicolumn{5}{c}{Unquenched mass: 10896.2 MeV}\\
\multicolumn{5}{c}{Exp. mass: 10885.2 MeV}\\
\hline
\end{tabular}
\end{center}
\end{table}

From the Table \ref{results}, we can find that $B\bar{B} (0,0,0)$, a negative mass shift of -4.87 MeV is obtained. The	vector-vector channel $B^{*}\bar{B}^{*} (1,1,0)$ with total spin $S=0$ yields a small mass shift of -2.39 MeV. For $B\bar{B}^{*} (0,1,1)$, an attractive shift of -1.96 MeV arises. The channel $B^{*}\bar{B} (1,0,1)$ is isospin-related to the previous one and produces an identical mass shift of -1.96 MeV. Exchanging the spin assignment of the two heavy mesons does not modify either the spin-flip strength or the phase-space suppression, so the quantitative impact remains the same. For the channel $B^{*}\bar{B}^{*} (1,1,1)$, the total spin $S=1$ vector-vector interaction produces a vanishing net contribution ($\Delta M \simeq 0$). For the channel $B^{*}\bar{B}^{*} (1,1,2)$, the total spin $S=2$ $D$-wave component contributes a positive mass shift +8.0 MeV. 
As all the $\Upsilon(nS)$ with $1^{--}$ states are taken into account in the calculation, the observed positive mass shift may result from the influence of the $4S$ state or potentially more suitable nearby states.

In the case of strange meson coupling channels, an attraction of -2.99 MeV is found for $B_s\bar{B}_s (0,0,0)$ channel. For the $B_s^*\bar{B}_s^* (1,1,0)$ channel, with total spin $S=0$, the strange vector-vector channel yields -2.4 MeV. A pronounced shift of -8.22 MeV emerges for $B_s\bar{B}_s^* (0,1,1)$. Together with $B_s^*\bar{B}_s (1,0,1)$ channel, the modes account for nearly 52 \% of the total mass shift. For the $B_s^*\bar{B}_s^*$ channel, the $D$-wave component imparts the relatively largest single-channel attraction of -6.4 MeV.

Summing all channels yields a cumulative shift of $\Delta M = -31.4$ MeV, lowering the bare quark-model mass from 10927.6 MeV to 10896.2 MeV for the $\Upsilon(5S)$ state. The residual 11 MeV difference from the PDG average of 10885.2 MeV \cite{ParticleDataGroup:2024cfk} underscores the quantitative adequacy of the coupled-channel description and indicates that the $\Upsilon(5S)$ is a strongly mixed $b\bar{b}-B^{(*)}_{(s)}B^{(*)}_{(s)}$ state rather than a pure bottomonium excitation. 

What's more, we found that when we consider the effects of the coupled-channels, the mass of the high-lying excitation $\Upsilon(5S)$ is reduced to the 10896.2 MeV, which is very consistent with the experimental value 10885.2 MeV. So in our calculations, the state $\Upsilon(5S)$ is not a pure $b\bar{b}$ state, but a mixture state of $b\bar{b}$ and four-quark components.

Lastly, in the unquenched quark model framework, we can not lower the mass of $\Upsilon(10860)$ to that of $\Upsilon(10753)$. So in our calculations, we also 
support the idea that the $\Upsilon(10860)$ and $\Upsilon(10753)$ is not the same state. Regarding the $\Upsilon(10753)$ state, its internal structure and nature—whether it is a conventional $b\bar{b}$ state or an exotic state, requirs further investigation in future studies. Additionally, the coupled-channel effects on all other highly excited $b\bar{b}$ states involve substantial computational complexity. This will be the focus of our subsequent work.

%%%%%%%%%%%%%%%%%%%%%%%%%%%%%%%%%%%%%%%%%%%%%%%%%%%%%%%%%%%%%

\section{Summary}
\label{epilogue}
In this manuscript, the mass of the $\Upsilon(5S)$ is calculated taking into account coupling to the pairs of lowest $B$ and $B_s$ pairs in the unquenched quark model, to compare the $\Upsilon(10753)$ state experimentally. 
To minimize the error from the calculation, a powerful method for dealing with few-body systems (GEM) was used. In our work, the angular momentum of the two mesons taken zero, and the relative motion between the two mesons denotes to $P$ wave for $J^{PC}=1^{--}$ states.
The transition operator of the $^3P_0$ model is required to relate the valence part to the four-quark components. We demonstrated the mass shifts of the $\Upsilon(5S)$ with the modified version of the transition operator. 

We obtained the unquenched mass of the $\Upsilon(5S)$ to be 10896.2 MeV, which is very close to the experimental values 10885.2 MeV of $\Upsilon(10860)$ state, but is higher than the mass of the $\Upsilon(10753)$ state. And in our work, we regard the $\Upsilon(10860)$ and the $\Upsilon(10753)$ state as the different states.
As for the other $\Upsilon(nS)$ with $1^{--}$ states, the coupled-channels effects may also play an important role, which will be our next study in the future.

%%%%%%%%%%%%%%%%%%%%%%%%%%%%%%%%%%%%%%%%%%%%%%%%%%%%%%%%%%%%%%


\begin{thebibliography}{10}
\bibitem{Liu:2013waa}
X.~Liu,
%``An overview of $XYZ$ new particles,''
Chin. Sci. Bull. \textbf{59}, 3815-3830 (2014)
doi:10.1007/s11434-014-0407-2
[arXiv:1312.7408 [hep-ph]].

\bibitem{Chen:2016qju}
H.~X.~Chen, W.~Chen, X.~Liu and S.~L.~Zhu,
%``The hidden-charm pentaquark and tetraquark states,''
Phys. Rept. \textbf{639}, 1-121 (2016)
doi:10.1016/j.physrep.2016.05.004
[arXiv:1601.02092 [hep-ph]].

\bibitem{Liu:2019zoy}
Y.~R.~Liu, H.~X.~Chen, W.~Chen, X.~Liu and S.~L.~Zhu,
%``Pentaquark and Tetraquark states,''
Prog. Part. Nucl. Phys. \textbf{107}, 237-320 (2019)
doi:10.1016/j.ppnp.2019.04.003
[arXiv:1903.11976 [hep-ph]].

\bibitem{Guo:2017jvc}
F.~K.~Guo, C.~Hanhart, U.~G.~Mei\ss{}ner, Q.~Wang, Q.~Zhao and B.~S.~Zou,
%``Hadronic molecules,''
Rev. Mod. Phys. \textbf{90}, no.1, 015004 (2018)
[erratum: Rev. Mod. Phys. \textbf{94}, no.2, 029901 (2022)]
doi:10.1103/RevModPhys.90.015004
[arXiv:1705.00141 [hep-ph]].

\bibitem{Brambilla:2019esw}
N.~Brambilla, S.~Eidelman, C.~Hanhart, A.~Nefediev, C.~P.~Shen, C.~E.~Thomas, A.~Vairo and C.~Z.~Yuan,
%``The $XYZ$ states: experimental and theoretical status and perspectives,''
Phys. Rept. \textbf{873}, 1-154 (2020)
doi:10.1016/j.physrep.2020.05.001
[arXiv:1907.07583 [hep-ex]].

\bibitem{Wang:2021aql}
F.~L.~Wang, X.~D.~Yang, R.~Chen and X.~Liu,
%``Correlation of the hidden-charm molecular tetraquarks and the charmoniumlike structures existing in the $B\to XYZ+K$ process,''
Phys. Rev. D \textbf{104}, no.9, 094010 (2021)
doi:10.1103/PhysRevD.104.094010
[arXiv:2103.04698 [hep-ph]].

%\bibitem{Belle:2019cbt}
%R.~Mizuk \textit{et al.} [Belle],
%``Observation of a new structure near 10.75 GeV in the energy dependence of the e$^{+}$e$^{?}$\textrightarrow{} \Upsilon{} (nS)\ensuremath{\pi}$^{+}$\ensuremath{\pi}$^{?}$ (n = 1, 2, 3) cross sections,''
%JHEP \textbf{10}, 220 (2019)
%doi:10.1007/JHEP10(2019)220
%[arXiv:1905.05521 [hep-ex]].

\bibitem{Belle-II:2022xdi}
I.~Adachi \textit{et al.} [Belle-II],
%``Observation of e+e-\textrightarrow{}\ensuremath{\omega}\ensuremath{\chi}bJ(1P) and Search for Xb\textrightarrow{}\ensuremath{\omega}\Upsilon{}(1S) at s near 10.75~GeV,''
Phys. Rev. Lett. \textbf{130}, no.9, 091902 (2023)
doi:10.1103/PhysRevLett.130.091902
[arXiv:2208.13189 [hep-ex]].

\bibitem{Chen:2019uzm}
B.~Chen, A.~Zhang and J.~He,
%``Bottomonium spectrum in the relativistic flux tube model,''
Phys. Rev. D \textbf{101}, no.1, 014020 (2020)
doi:10.1103/PhysRevD.101.014020
[arXiv:1910.06065 [hep-ph]].

\bibitem{Liang:2019geg}
W.~H.~Liang, N.~Ikeno and E.~Oset,
%``$\Upsilon(nl)$ decay into $ B^{(*)} \bar B^{(*)}$,''
Phys. Lett. B \textbf{803}, 135340 (2020)
doi:10.1016/j.physletb.2020.135340
[arXiv:1912.03053 [hep-ph]].

\bibitem{ATLAS:2020yzc}
G.~Aad \textit{et al.} [ATLAS],
%``Search for dark matter produced in association with a single top quark in $\sqrt{s}=13$~TeV $pp$ collisions with the ATLAS detector,''
Eur. Phys. J. C \textbf{81}, 860 (2021)
doi:10.1140/epjc/s10052-021-09566-y
[arXiv:2011.09308 [hep-ex]].

\bibitem{Giron:2020qpb}
J.~F.~Giron and R.~F.~Lebed,
%``Spectrum of the hidden-bottom and the hidden-charm-strange exotics in the dynamical diquark model,''
Phys. Rev. D \textbf{102}, no.1, 014036 (2020)
doi:10.1103/PhysRevD.102.014036
[arXiv:2005.07100 [hep-ph]].

\bibitem{vanBeveren:2020eis}
E.~van Beveren and G.~Rupp,
%``Modern meson spectroscopy: the fundamental role of unitarity,''
Prog. Part. Nucl. Phys. \textbf{117}, 103845 (2021)
doi:10.1016/j.ppnp.2020.103845
[arXiv:2012.03693 [hep-ph]].

\bibitem{Li:2021jjt}
Y.~S.~Li, Z.~Y.~Bai, Q.~Huang and X.~Liu,
%``Hidden-bottom hadronic decays of \Upsilon{}(10753) with a \ensuremath{\eta}(') or \ensuremath{\omega} emission,''
Phys. Rev. D \textbf{104}, no.3, 034036 (2021)
doi:10.1103/PhysRevD.104.034036
[arXiv:2106.14123 [hep-ph]].

\bibitem{Bai:2022cfz}
Z.~Y.~Bai, Y.~S.~Li, Q.~Huang, X.~Liu and T.~Matsuki,
%``\Upsilon{}(10753)\textrightarrow{}\Upsilon{}(nS)\ensuremath{\pi}+\ensuremath{\pi}- decays induced by hadronic loop mechanism,''
Phys. Rev. D \textbf{105}, no.7, 074007 (2022)
doi:10.1103/PhysRevD.105.074007
[arXiv:2201.12715 [hep-ph]].

\bibitem{TarrusCastella:2021pld}
J.~Tarr\'us Castell\`a and E.~Passemar,
%``Exotic to standard bottomonium transitions,''
Phys. Rev. D \textbf{104}, no.3, 034019 (2021)
doi:10.1103/PhysRevD.104.034019
[arXiv:2104.03975 [hep-ph]].

\bibitem{Ali:2019okl}
A.~Ali, L.~Maiani, A.~Y.~Parkhomenko and W.~Wang,
%``Interpretation of $Y_b (10753)$ as a tetraquark and its production mechanism,''
Phys. Lett. B \textbf{802}, 135217 (2020)
doi:10.1016/j.physletb.2020.135217
[arXiv:1910.07671 [hep-ph]].

\bibitem{Wang:2019veq}
Z.~G.~Wang,
%``Vector hidden-bottom tetraquark candidate: $Y(10750)$,''
Chin. Phys. C \textbf{43}, no.12, 123102 (2019)
doi:10.1088/1674-1137/43/12/123102
[arXiv:1905.06610 [hep-ph]].

\bibitem{Bicudo:2020qhp}
P.~Bicudo, N.~Cardoso, L.~Mueller and M.~Wagner,
%``Computation of the quarkonium and meson-meson composition of the $\Upsilon(nS)$  states and of the new $\Upsilon(10753)$ Belle resonance from lattice QCD static potentials,''
Phys. Rev. D \textbf{103}, no.7, 074507 (2021)
doi:10.1103/PhysRevD.103.074507
[arXiv:2008.05605 [hep-lat]].

\bibitem{Bicudo:2022ihz}
P.~Bicudo, N.~Cardoso, L.~Mueller and M.~Wagner,
%``Study of I=0 bottomonium bound states and resonances in S, P, D, and F waves with lattice QCD static-static-light-light potentials,''
Phys. Rev. D \textbf{107}, no.9, 094515 (2023)
doi:10.1103/PhysRevD.107.094515

\bibitem{CLEO:1984vfn}
D.~Besson \textit{et al.} [CLEO],
%``Observation of New Structure in the e+ e- Annihilation Cross-Section Above B anti-B Threshold,''
Phys. Rev. Lett. \textbf{54}, 381 (1985)

\bibitem{ParticleDataGroup:2024cfk}
S.~Navas \textit{et al.} [Particle Data Group],
%``Review of particle physics,''
Phys. Rev. D \textbf{110}, no.3, 030001 (2024)
doi:10.1103/PhysRevD.110.030001

\bibitem{Vijande:2004he}
J.~Vijande, F.~Fernandez and A.~Valcarce,
%``Constituent quark model study of the meson spectra,''
J. Phys. G \textbf{31}, 481 (2005)
doi:10.1088/0954-3899/31/5/017
[arXiv:hep-ph/0411299 [hep-ph]].

\bibitem{Hiyama:2003cu}
E.~Hiyama, Y.~Kino and M.~Kamimura,
%``Gaussian expansion method for few-body systems,''
Prog. Part. Nucl. Phys. \textbf{51}, 223-307 (2003)
doi:10.1016/S0146-6410(03)90015-9

\bibitem{Tang:1978zz}
Y.~C.~Tang, M.~Lemere and D.~R.~Thompson,
%``Resonating-group method for nuclear many-body problems,''
Phys. Rept. \textbf{47}, 167-223 (1978)
doi:10.1016/0370-1573(78)90175-8

\bibitem{Wheeler:1937zza}
J.~A.~Wheeler,
%``Molecular Viewpoints in Nuclear Structure,''
Phys. Rev. \textbf{52}, 1083-1106 (1937)
doi:10.1103/PhysRev.52.1083

\bibitem{Micu:1968mk}
L.~Micu,
%``Decay rates of meson resonances in a quark model,''
Nucl. Phys. B \textbf{10}, 521-526 (1969)
doi:10.1016/0550-3213(69)90039-X

\bibitem{Barnes:2007xu}
T.~Barnes and E.~S.~Swanson,
%``Hadron loops: General theorems and application to charmonium,''
Phys. Rev. C \textbf{77}, 055206 (2008)
doi:10.1103/PhysRevC.77.055206
[arXiv:0711.2080 [hep-ph]].

\bibitem{Ferretti:2018tco}
J.~Ferretti and E.~Santopinto,
%``Threshold corrections of $\chi_c$ (2 P ) and $\chi_b$ (3 P ) states and J /$\psi \rho$ and J /$\psi \omega$ transitions of the $\chi$ (3872) in a coupled-channel model,''
Phys. Lett. B \textbf{789}, 550-555 (2019)
doi:10.1016/j.physletb.2018.12.052
[arXiv:1806.02489 [hep-ph]].

\bibitem{Chen:2017mug}
X.~Chen, J.~Ping, C.~D.~Roberts and J.~Segovia,
%``Light-meson masses in an unquenched quark model,''
Phys. Rev. D \textbf{97}, no.9, 094016 (2018)
doi:10.1103/PhysRevD.97.094016
[arXiv:1712.04457 [nucl-th]].

\bibitem{Chen:2024ukv}
X.~Chen and Y.~Tan,
%``Light quarkonium and charmonium mass shifts in an unquenched quark model*,''
Chin. Phys. C \textbf{48}, no.9, 093107 (2024)
doi:10.1088/1674-1137/ad53bd
[arXiv:2406.00957 [hep-ph]].

\bibitem{Chen:2023wfp}
X.~Chen, Y.~Tan and Y.~Yang,
%``Charmonium mass shifts in an unquenched quark model,''
Eur. Phys. J. Plus \textbf{138}, no.7, 653 (2023)
doi:10.1140/epjp/s13360-023-04181-0
[arXiv:2301.12388 [hep-ph]].

\bibitem{Machleidt:1987hj}
R.~Machleidt, K.~Holinde and C.~Elster,
%``The Bonn Meson Exchange Model for the Nucleon Nucleon Interaction,''
Phys. Rept. \textbf{149}, 1-89 (1987)
doi:10.1016/S0370-1573(87)80002-9

\bibitem{Valcarce:2005em}
A.~Valcarce, H.~Garcilazo, F.~Fernandez and P.~Gonzalez,
%``Quark-model study of few-baryon systems,''
Rept. Prog. Phys. \textbf{68}, 965-1042 (2005)
doi:10.1088/0034-4885/68/5/R01

\bibitem{LeYaouanc:1972vsx}
A.~Le Yaouanc, L.~Oliver, O.~Pene and J.~C.~Raynal,
%``Naive quark pair creation model of strong interaction vertices,''
Phys. Rev. D \textbf{8}, 2223-2234 (1973)
doi:10.1103/PhysRevD.8.2223

\bibitem{LeYaouanc:1973ldf}
A.~Le Yaouanc, L.~Oliver, O.~Pene and J.~C.~Raynal,
%``Naive quark pair creation model and baryon decays,''
Phys. Rev. D \textbf{9}, 1415-1419 (1974)
doi:10.1103/PhysRevD.9.1415

\bibitem{Roberts:1992js}
W.~Roberts and B.~Silvestre- Brac,
%``General method of calculation of any hadronic decay in the p wave triplet model,''
Acta Phys. Austriaca \textbf{11}, 171-193 (1992)

\bibitem{Capstick:1993kb}
S.~Capstick and W.~Roberts,
%``Quasi two-body decays of nonstrange baryons,''
Phys. Rev. D \textbf{49}, 4570-4586 (1994)
doi:10.1103/PhysRevD.49.4570

\bibitem{Page:1995rh}
P.~R.~Page,
%``Excited charmonium decays by flux tube breaking and the psi-prime anomaly at CDF,''
Nucl. Phys. B \textbf{446}, 189-210 (1995)
doi:10.1016/0550-3213(95)00171-N

\bibitem{Ackleh:1996yt}
E.~S.~Ackleh, T.~Barnes and E.~S.~Swanson,
%``On the mechanism of open flavor strong decays,''
Phys. Rev. D \textbf{54}, 6811-6829 (1996)
doi:10.1103/PhysRevD.54.6811

\bibitem{Segovia:2012cd}
J.~Segovia, D.~R.~Entem and F.~Fern\'andez,
%``Scaling of the $^3P_0$ Strength in Heavy Meson Strong Decays,''
Phys. Lett. B \textbf{715}, 322-327 (2012)
doi:10.1016/j.physletb.2012.08.005

\bibitem{LeYaouanc:1977gm}
A.~Le Yaouanc, L.~Oliver, O.~Pene and J.~C.~Raynal,
%``Why Is psi-prime-prime-prime (4.414) SO Narrow?,''
Phys. Lett. B \textbf{72}, 57-61 (1977)
doi:10.1016/0370-2693(77)90062-4

\end{thebibliography}
\end{document}